\documentclass[conference]{IEEEtran}
\IEEEoverridecommandlockouts

\usepackage{cite}
\usepackage{amsmath,amssymb,amsfonts}
\usepackage{algorithmic}
\usepackage{graphicx}
\usepackage{textcomp}
\usepackage{xcolor}
\def\BibTeX{{\rm B\kern-.05em{\sc i\kern-.025em b}\kern-.08em
    T\kern-.1667em\lower.7ex\hbox{E}\kern-.125emX}}
\begin{document}

\title{\fontsize{22}{24}\selectfont \textbf{L-SPINE:} A Low-Precision SIMD Spiking Neural Compute Engine for Resource-efficient Edge Inference\\
\thanks{This work was supported partially by the Dept of Science and Technology (DST), Govt of India, for the INSPIRE PhD fellowship.}

}

\author{\IEEEauthorblockN{Sonu Kumar}
\IEEEauthorblockA{\textit{Centre for Advanced Electronics} \\
\textit{Indian Institute of Technology Indore}\\
Simrol 453552, Indore, India\\
phd2101191002@iiti.ac.in}
\and
\IEEEauthorblockN{Mukul Lokhande}
\IEEEauthorblockA{\textit{Dept. of Electrical Engineering} \\
\textit{Indian Institute of Technology Indore}\\
Simrol 453552, Indore, India\\
phd2201102020@iiti.ac.in}
\and
\IEEEauthorblockN{ Santosh Kumar Vishvakarma}
\IEEEauthorblockA{\textit{Dept. of Electrical Engineering} \\
\textit{Indian Institute of Technology Indore}\\
Simrol 453552, Indore, India\\
skvishvakarma@iiti.ac.in}
}

\maketitle

\begin{abstract}

Spiking Neural Networks (SNNs) offer a promising solution for energy-efficient edge intelligence; however, their hardware deployment is constrained by memory overhead, inefficient scaling operations, and limited parallelism. This work proposes L-SPINE, a low-precision SIMD-enabled spiking neural compute engine for efficient edge inference. The architecture features a unified multi-precision datapath supporting 2-bit, 4-bit, and 8-bit operations, leveraging a multiplier-less shift-add model for neuron dynamics and synaptic accumulation. Implemented on an AMD VC707 FPGA, the proposed neuron requires only 459 LUTs and 408 FFs, achieving a critical delay of 0.39 ns and 4.2 mW power. At the system level, L-SPINE achieves 46.37K LUTs, 30.4K FFs, 2.38 ms latency, and 0.54 W power. Compared to CPU and GPU platforms, it reduces inference latency from seconds to milliseconds, achieving an up to three orders-of-magnitude improvement in energy efficiency. Quantisation analysis shows that INT2/INT4 configurations significantly reduce memory footprint with minimal accuracy loss. These results establish L-SPINE as a scalable and efficient solution for real-time edge SNN deployment.

\end{abstract}


\section{Introduction}

Spiking Neural Networks (SNNs) have emerged as a promising paradigm for energy-efficient edge intelligence by leveraging sparse event-driven computation and temporal information processing \cite{roy2019spike, rathi2023survey, li2024brain}. Unlike conventional artificial neural networks (ANNs), SNNs operate using binary spike trains, significantly reducing computational complexity and power consumption, making them well-suited for resource-constrained edge platforms and neuromorphic systems \cite{Nature1, Nature2}. Recent advances in quantization-aware training methodologies, including backpropagation-through-time (BPTT), surrogate gradient learning, and sparsity-aware optimisation have substantially improved the accuracy of deep SNNs, enabling their deployment in complex vision, sensing, and edge intelligence applications \cite{rathi2021diet, lee2018deepstdp, rathi2018stdpprune, yin2022sata, Nature4, Nature3, 11000281, 11016825}.

Despite these advances, efficient hardware deployment of SNNs remains a critical challenge. First, the temporal nature of spike-based computation requires repeated storage and updates of membrane potentials across timesteps, leading to significant memory overhead and increased data movement. Second, existing quantisation and compression techniques, such as STBP-based quantisation and ADMM-based optimization, primarily focus on reducing weight precision while overlooking system-level inefficiencies arising from scaling operations and temporal dataflow \cite{STBP, ADMM, QuantMAC}. Recent works such as MINT \cite{10473825} and bit-shifting based integer inference approaches \cite{song2025bitshifting} have attempted to eliminate multiplier overheads; however, they do not fully exploit hardware-level parallelism or provide a unified framework for multi-precision SNN execution.

\begin{figure}[!t]
    \centering
    \includegraphics[width=0.875\columnwidth]{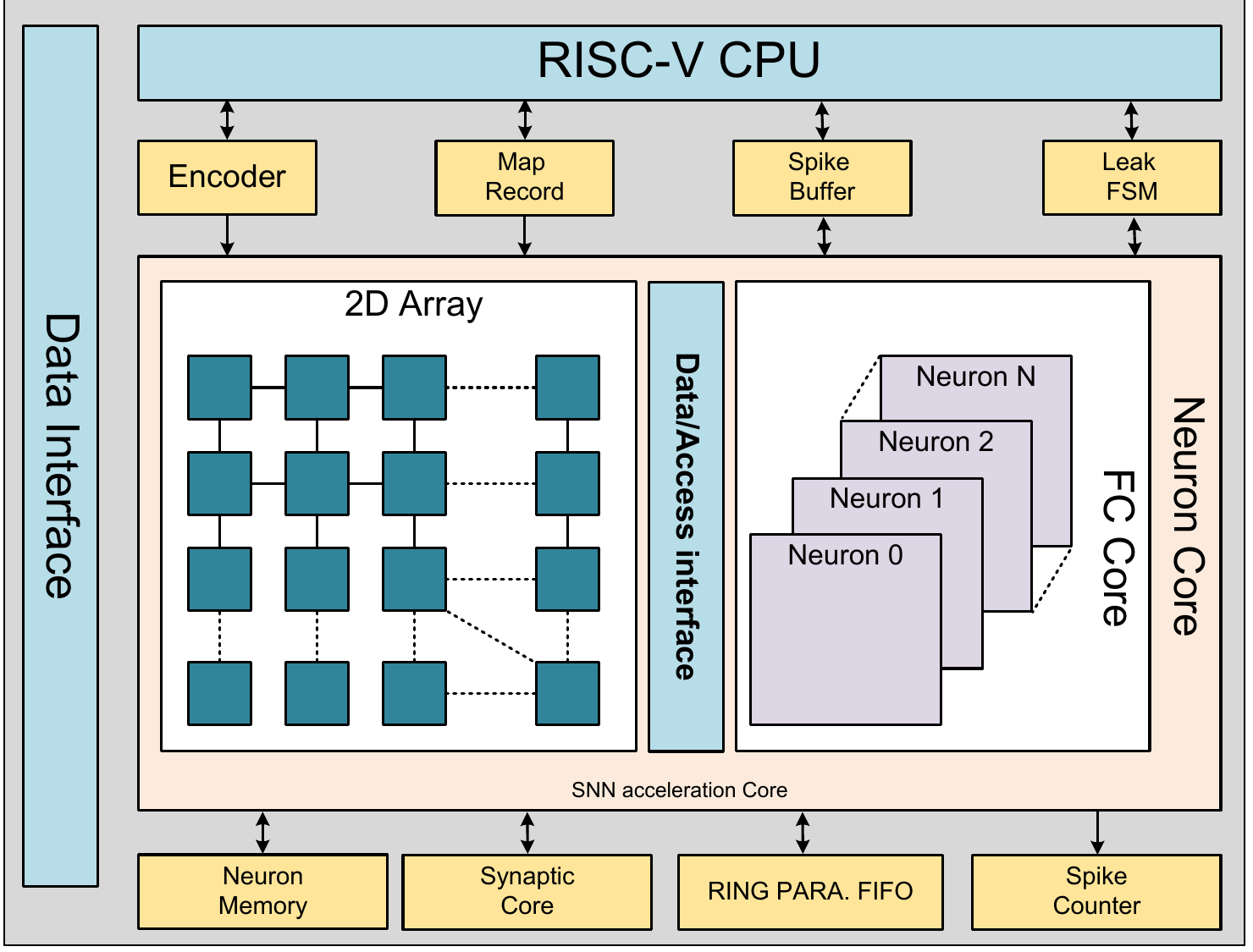}
    \caption{System-level architecture of the proposed L-SPINE accelerator integrating pico-rv32 RISC-V, spike encoding modules, and a 2D SIMD-enabled neuron processing array for efficient SNN inference.}
    \label{fig:arch}
\end{figure}

\begin{figure*}[!t]
    \centering
    \includegraphics[width=\textwidth]{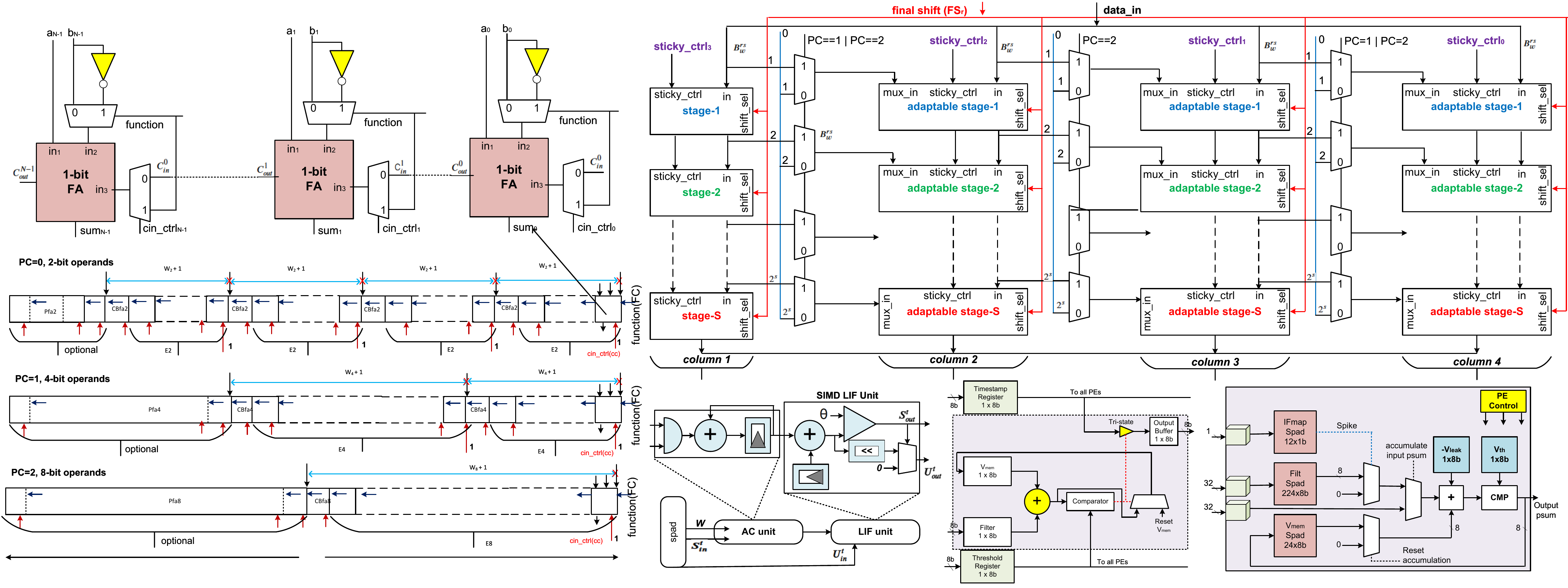}
    \caption{Detailed datapath for Proposed SIMD-enabled multi-precision compute engine supporting configurable 16x 2-bit, 4x 4-bit, and 1x 8-bit operations using a reconfigurable shift-add logic integrated in LIF neuron computation.}
    \label{fig:pe}
\end{figure*}

On the hardware front, prior SNN accelerators have explored neuron-level optimizations and dataflow strategies using CORDIC-based implementations, lookup-table approximations, and systolic architectures \cite{HH-TCASI'21, CORDIC_Izhikevich_TBioCAS'22, Izhikevich_NEWCAS'21, TCAS-I'19}. While these designs improve computational efficiency, they often suffer from limited scalability, lack of precision adaptability, and inefficient utilization of parallel compute resources. Several works have demonstrated FPGA and ASIC-based neuromorphic accelerators targeting low-power applications \cite{TCAD'23, TCAS-I'22, Neuromorphic_accl}, incorporating biologically inspired neuron models such as Leaky Integrate-and-Fire (LIF), Hodgkin–Huxley (H\&H), and Izhikevich neurons. These implementations employ techniques including CORDIC-based computation, multiplier-less arithmetic, piecewise linear (PWL) approximation, and lookup-table-based designs to improve hardware efficiency \cite{TCASI22, TCAS-I_24_Gomar, TCAS-II_19_Hayati, TCAS-II_19_Farsa, ACMTR}.

Furthermore, recent SIMD-enabled compute engines for deep neural networks (DNNs) have demonstrated significant improvements in throughput and resource efficiency via multi-precision execution and hardware reuse \cite{Flex-PE, DNN_Retro'25, LPRE}. However, their applicability to SNNs is limited due to the temporal dynamics and event-driven nature of spike-based computation. Existing designs lack a unified architecture that supports multi-precision execution, efficient neuron dynamics, and SIMD parallelism within a single datapath, underscoring the need for a scalable, hardware-efficient SNN accelerator for parallel low-bit inference.

To address these challenges, we propose \textbf{L-SPINE}, a low-precision SIMD-enabled spiking neural compute engine for resource-efficient edge inference. The architecture features a unified reconfigurable datapath supporting 2-, 4-, and 8-bit operations, leveraging a multiplier-less shift-add model for neuron dynamics and synaptic accumulation. Additionally, a tightly coupled system integrating a RISC-V controller, spike encoding modules, and a scalable 2D neuron array enables efficient dataflow and real-time deployment.

\noindent \textbf{The key contributions of this work are as follows:}
\begin{itemize}
\item A unified multi-precision SIMD compute engine enabling parallel execution of 2/4/8-bit SNN operations within a single datapath, supporting up to \textbf{16× (INT2), 4× (INT4), and 1× (INT8)} parallelism, thereby significantly improving throughput for low-bit inference.

\item A multiplier-less neuron computation framework based on shift-add operations, achieving a compact implementation with only \textbf{459 LUTs and 408 FFs}, \textbf{0.39 ns} critical delay, and \textbf{4.2 mW} power consumption.

\item A scalable system-level SNN accelerator integrating a RISC-V controller and 2D neuron array, achieving \textbf{2.38 ms} latency with only \textbf{0.54 W} power, enabling real-time edge inference.

\item Comprehensive evaluation demonstrating \textbf{orders-of-magnitude latency reduction} (seconds to milliseconds) and improved energy efficiency compared to CPU/GPU baselines, while maintaining competitive accuracy under INT2/INT4 quantisation.

\end{itemize}

\section{L-SPINE Architecture and Design Framework}

\subsection{System-Level Architecture}

Fig.~\ref{fig:arch} illustrates the overall architecture of the proposed \textbf{L-SPINE} compute engine, designed for efficient edge deployment of spiking neural networks (SNNs). The system integrates a tightly coupled \textit{RISC-V control unit} with a dedicated SNN acceleration core comprising a 2D processing array, neuron memory, synaptic computation units, and spike management modules. The input data is first processed through an encoder and mapped into spike representations, which are stored in the spike buffer. The RISC-V controller orchestrates data movement, scheduling, and synchronisation across the accelerator. A ring-based FIFO interface facilitates efficient data transfer between memory and compute units, minimizing access latency and bandwidth overhead. The SNN core consists of a scalable 2D array of neuron compute engines (NCEs), each of which performs synaptic accumulation and updates the neuron state. The architecture adopts a dataflow that exploits temporal reuse of membrane potentials and spatial reuse of synaptic weights, enabling efficient computation across multiple timesteps. Additionally, dedicated modules such as the leak FSM and spike counter support neuron dynamics and event-driven execution.

The proposed L-SPINE architecture is driven by several key design principles. It enables multi-precision SIMD execution, allowing dynamic adaptation to different quantisation levels (INT2–8) to improve throughput for low-bitwidth SNN inference. The architecture employs a multiplier-less computation model based on shift-add operations for membrane updates and synaptic accumulation, significantly reducing hardware complexity. A unified datapath design supports multiple precision modes within a single NCE, avoiding redundant hardware resources. Additionally, dataflow optimisation exploits temporal reuse of membrane potentials and spatial reuse of weights to minimise memory bandwidth requirements.

\subsection{SIMD Multi-Precision compute engine}

Fig.~\ref{fig:pe} presents the proposed multi-precision SIMD-enabled compute engine (NCE), which forms the computational backbone of the L-SPINE architecture. The NCE supports configurable precision modes, enabling 2-, 4-, and 8-bit operations within a unified hardware datapath. The MAC computation is realised using a hierarchy of 1-bit full adders arranged in a bit-serial and parallel hybrid configuration. Depending on the selected precision control (PC), the datapath dynamically reconfigures to perform multiple low-bitwidth operations in parallel, effectively enabling SIMD-style execution. This allows higher throughput for low-precision workloads while maintaining flexibility for higher precision computations. The accumulation (AC) unit performs spike-driven synaptic integration using integer arithmetic, where input spikes (binary events) are accumulated with quantized weights. The membrane potential update is implemented using shift-based operations to model leakage behaviour, eliminating the need for multipliers. A comparator-based thresholding unit generates output spikes based on the neuron firing condition, while a reset mechanism ensures proper temporal dynamics. The integration of the AC unit and LIF neuron model within a single NCE enables efficient temporal processing and reduces data movement overhead. Furthermore, local scratchpad memories store input feature maps, weights, and membrane potentials, ensuring high data locality and minimizing external memory accesses.

\section{Methodology \& Performance Evaluation}

\subsection{Design and Evaluation Flow}

\begin{figure}[!t]
    \centering
    \includegraphics[width=0.875\columnwidth]{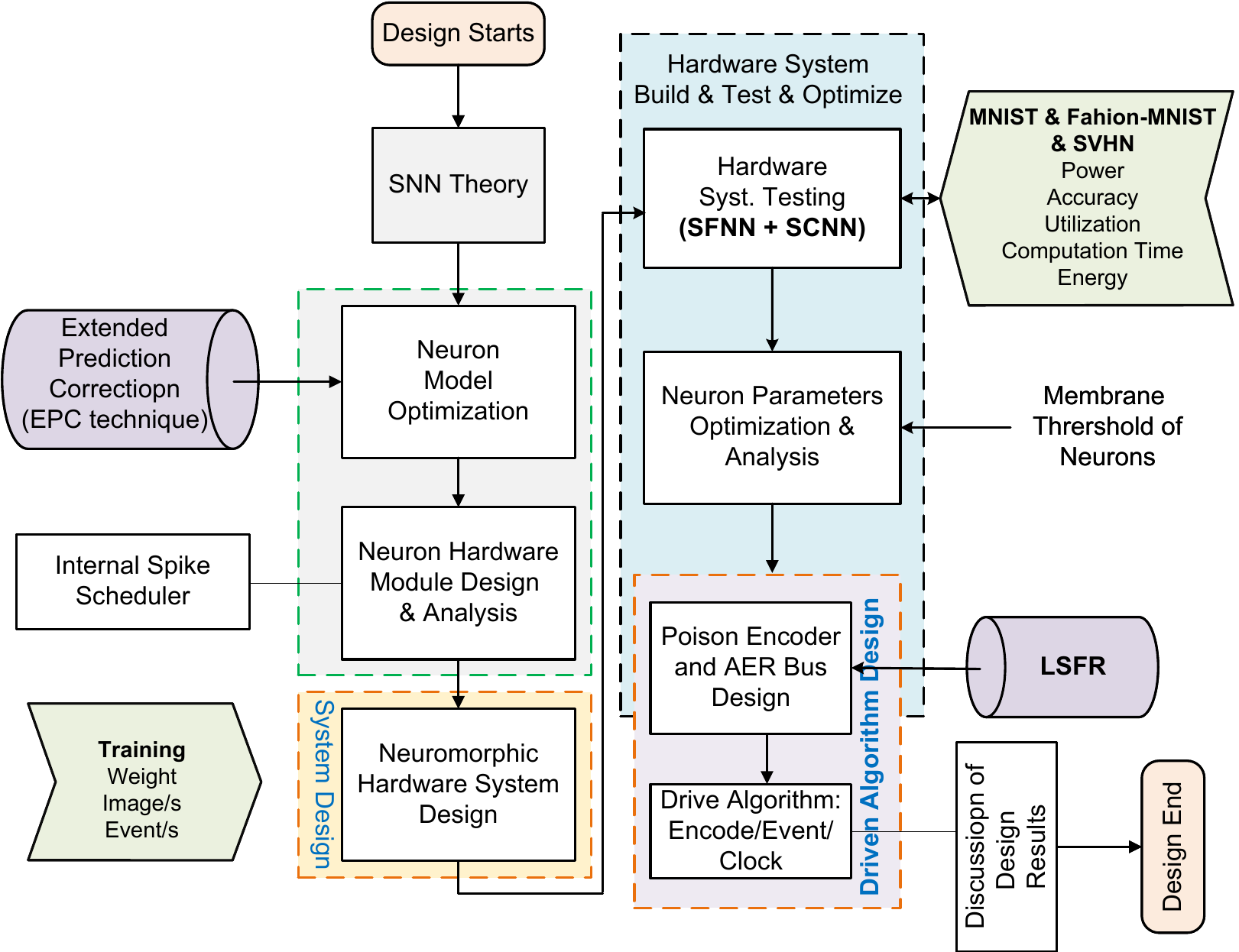}
    \caption{Design and evaluation flow of the proposed L-SPINE architecture, including SNN training, quantization, hardware mapping, and FPGA-based validation.}
    \label{fig:method}
\end{figure}

The overall design and evaluation flow of the proposed L-SPINE architecture is illustrated in Fig.~\ref{fig:method}. The workflow begins with training SNN models using standard backpropagation-based techniques, followed by post-training quantization to low-bit precision formats (2-bit, 4-bit, and 8-bit). The quantised weights and neuron parameters are then mapped onto the proposed L-SPINE engine. The hardware design is implemented using a parameterised RTL model, enabling configurable precision modes and scalable compute engine (PE) array configurations. Functional verification is carried out using cycle-accurate simulations, followed by FPGA synthesis and implementation on a AMD Virtex-7 VC707 platform. The evaluation framework includes accuracy analysis, memory footprint estimation, and hardware performance metrics such as latency, power, and resource utilisation.

\subsection{Quantization Analysis}

\begin{figure}[!t]
    \centering
    \includegraphics[width=\columnwidth]{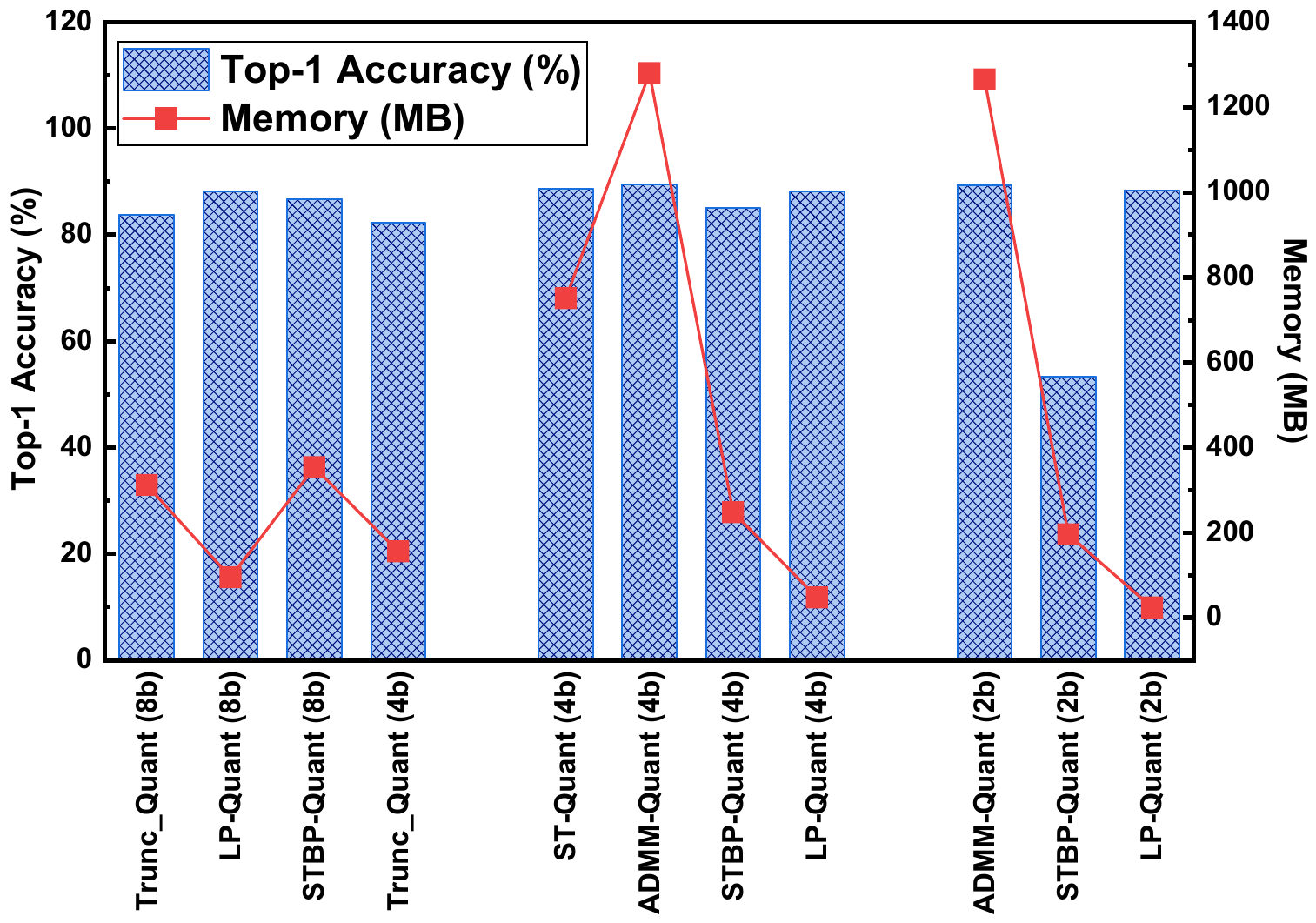}
    \caption{Comparison of accuracy and memory footprint with the state-of-the-art SNN quantisation, STBP\cite{STBP}, ADMM\cite{ADMM}, Trunc\cite{QuantMAC}.}
    \label{fig:quant_mem}
\end{figure}

\begin{figure}[!t]
    \centering
    \includegraphics[width=0.875\columnwidth]{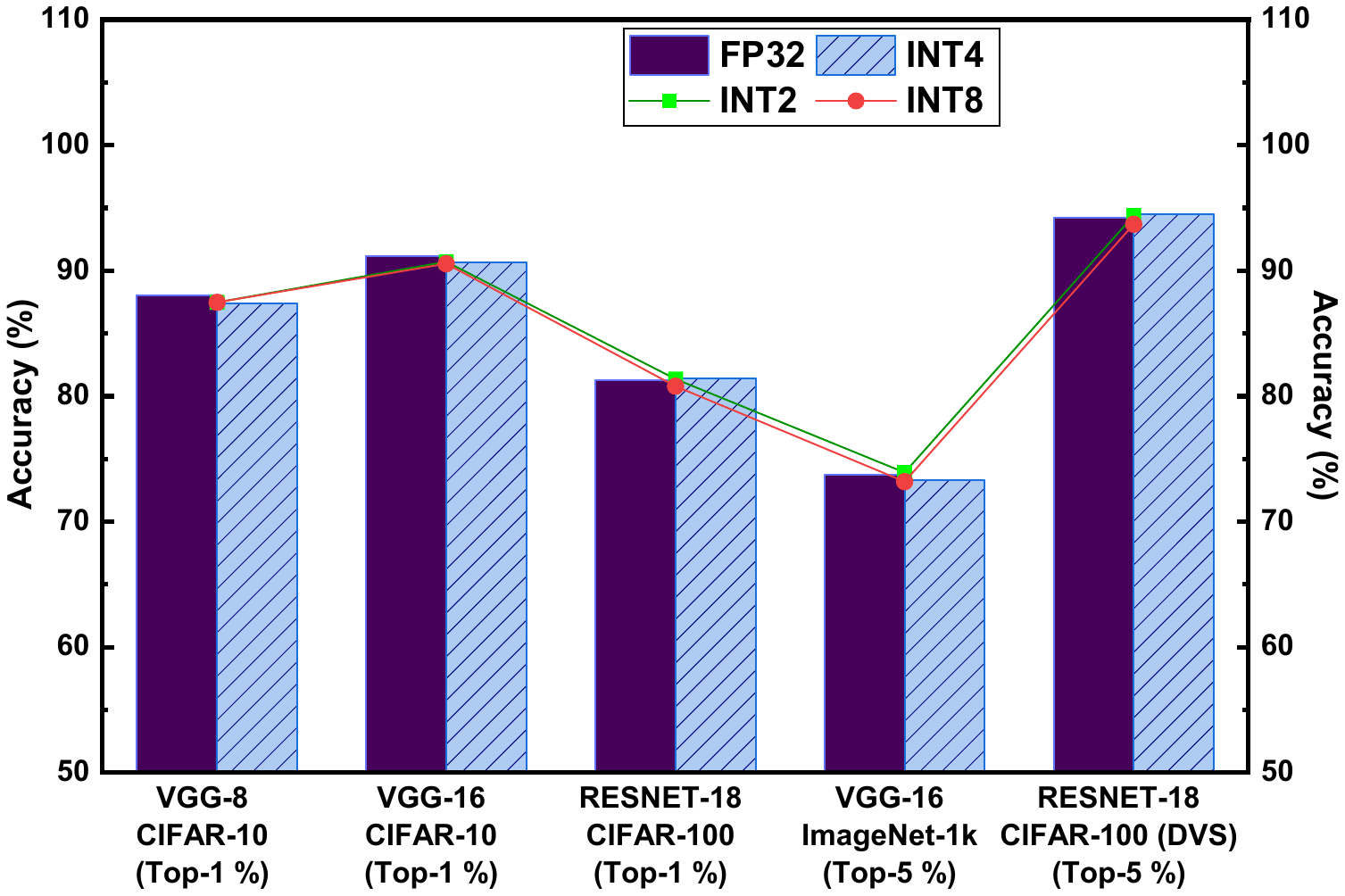}
    \caption{Impact of precision scaling on SNN accuracy across INT2, INT4, INT8, and FP32 configurations.}
    \label{fig:quant_acc}
\end{figure}

To evaluate the impact of quantization, we compare the proposed approach with state-of-the-art SNN quantization techniques including STBP~\cite{STBP}, ADMM~\cite{ADMM}, and Truncation-based methods~\cite{QuantMAC}. As shown in Fig.~\ref{fig:quant_mem}, the proposed approach achieves a favorable trade-off between accuracy and memory footprint across different precision levels. Specifically, aggressive low-bit quantisation (2-bit and 4-bit) significantly reduces memory requirements while maintaining competitive accuracy. This is achieved by efficiently mapping quantised parameters onto the SIMD datapath, minimising redundant storage and enabling a compact neuron state representation. Further, Fig.~\ref{fig:quant_acc} illustrates the effect of different precision levels on model accuracy. The results show that INT8 closely matches FP32 baseline performance, while INT4 and INT2 configurations exhibit graceful accuracy degradation, making them suitable for energy-constrained edge applications.

\subsection{FPGA-Based Hardware Evaluation}

The proposed neuron architecture is synthesised and evaluated on the VC707 FPGA, and compared with prior state-of-the-art neuron implementations, as summarised in Table~\ref{tab:neuron-fpga}. The proposed design achieves the lowest resource utilization and latency among all compared designs, requiring only 459 LUTs and 408 flip-flops, with a critical path delay of 0.39~ns and power consumption of 4.2~mW. Compared to recent designs such as TVLSI'26~\cite{Kumar2026ReLANCE} and TCAS-II'24~\cite{RPE}, the proposed architecture achieves significant reductions in area and delay, demonstrating the efficiency of the SIMD-based multi-precision datapath and multiplier-less computation model. At the system level, the complete accelerator is evaluated against prior SNN and neuromorphic accelerators, as shown in Table~\ref{tab:arch-fpga}. The proposed architecture achieves a latency of 2.38~ms with only 0.54~W power consumption, outperforming existing designs in both performance and energy efficiency.

\subsection{Energy and Performance Comparison}

To evaluate the efficiency of the proposed L-SPINE architecture, we compare its energy consumption with state-of-the-art SNN and neuromorphic accelerators. TCAD'23~\cite{TCAD'23} and TVLSI'26~\cite{Kumar2026ReLANCE} report energy of 1.12~J and 0.80~J, respectively. Specialized neuron implementations further reduce energy from 28.06~mJ~\cite{HH-TCASI'21} to 0.04~mJ~\cite{TCAS-I_21_Heidarpour}, with additional works reporting 5.04~mJ~\cite{CORDIC_Izhikevich_TBioCAS'22}, 2.96~mJ~\cite{TCAS-I'22}, 2.34~mJ~\cite{TVLSI25}, 1.19~mJ~\cite{NC20}, 0.99~mJ~\cite{Access22}, 0.19~mJ~\cite{TVLSI_24_Neil}, and 0.10~mJ~\cite{ISCAS_21_Lammie}. In comparison, L-SPINE achieves improved energy efficiency through low-precision SIMD execution and a multiplier-less datapath, where reduced precision (INT2/INT4/INT8) lowers switching activity and efficient data reuse minimizes overall power consumption.

We further compare the inference performance of the proposed design with CPU and GPU platforms. For \textbf{VGG-16}, CPU (Intel i7, INT8) requires 23.97 s at 125~W, and GPU (GTX 1050Ti, INT8) takes 10.15 s at 75~W. In contrast, the proposed accelerator achieves 4.83 ms (INT2) and 16.94 ms (INT8) with only 0.54~W. GPU floating-point execution shows higher latency: FP32 and FP16 require 40.4 s and 39.9 s, respectively, while FxP16 reduces latency to 31.16 ms. For \textbf{ResNet-18}, CPU requires 34.43 s and GPU takes 10.26 s, whereas the proposed design achieves 7.84 ms (INT2) and 16.84 ms (INT8). These results demonstrate algorithm–hardware efficiency through low-precision SIMD execution, achieving significant gains in latency, power, and energy efficiency for real-time edge AI applications.

\begin{table}[!t]
\centering
\caption{Comparison for Neurons' FPGA Resources utilization \cite{TCAS-I_22_Soleimani,TCAS-II_19_Hayati,TCAS-II_19_Farsa,Flex-PE,TCAS-I_24_Gomar}}
\label{tab:neuron-fpga}
\renewcommand{\arraystretch}{1.1}
\resizebox{\columnwidth}{!}{%
\begin{tabular}{|l|r|r|r|r|}
\hline
\textbf{Design}& \multicolumn{1}{c|}{\textbf{LUTs}} & \multicolumn{1}{c|}{\textbf{FFs}} & \textbf{Delay (ns)} & \textbf{Power (mW)} \\ \hline
TVLSI'26\cite{Kumar2026ReLANCE} & 1770 & 862 & 1.41 & 8.9 \\ \hline
TCAS-II'24\cite{RPE} & 8054 & 1718 & 4.62 & 22.5 \\\hline
MP-RPE\cite{RPE} & 8065 & 1072 & 5.56 & 21.8 \\\hline
Iterative CORDIC H\&H\cite{HH-TCASI'21} & 2344 & 460 & 5.00 & 11.6 \\ \hline
PWL H\&H\cite{HH-TCASI'21} & 29130 & 25430 & 39.06 & 85.0 \\ \hline
Parallel CORDIC H\&H\cite{HH-TCASI'21} & 86032 & 50228 & 15.78 & 140.0 \\ \hline
Multiplier-less H\&H\cite{PWL_HH_Frontiers'14} & 5660 & 2840 & 11.77 & 18.5 \\ \hline
RAM H\&H\cite{PWL_HH_Frontiers'14} & 4735 & 1552 & 10.00 & 15.2 \\ \hline
CORDIC Izhikevich\cite{CORDIC_Izhikevich_TBioCAS'22} & 986 & 264 & 2.16 & 10.7 \\ \hline
TCAS-I'19\cite{TCAS-I'19} & 818 & 211 & 3.2 & 14.9\\ \hline
TCAS-I'22\cite{TCASI22} & 617 & 493 & 0.43 & 4.7 \\ \hline
\textbf{Proposed} & \textbf{459} & \textbf{408} & \textbf{0.39} & \textbf{4.2} \\
\hline
\end{tabular}}
\end{table}

\begin{table}[!t]
\centering
\caption{Hardware resource comparison with SoTA accelerators (VC707 FPGA)\cite{TVLSI_24_Neil,ISCAS_21_Lammie,TCAS-I_21_Heidarpour}}
\label{tab:arch-fpga}
\renewcommand{\arraystretch}{1.2}
\resizebox{\columnwidth}{!}{%
\begin{tabular}{|l|r|r|r|r|}
\hline
\textbf{Design} & \multicolumn{1}{c|}{\textbf{LUTs (K)}} & \multicolumn{1}{c|}{\textbf{FFs (K)}} & \multicolumn{1}{c|}{\textbf{Latency (ms)}} & \textbf{Power (W)} \\ \hline
TVLSI'26\cite{Kumar2026ReLANCE} & 118.6 & 57.8 & 5.04 & 1.85 \\ \hline
TRETS'23 \cite{DNN_Retro'25} & 115.0 & 115.0 & 21.46 & 2.10 \\ \hline
TCAD'23\cite{TCAD'23} & 170.4 & 113.2 & 7.38 & 2.40 \\ \hline
Iterative CORDIC H\&H\cite{HH-TCASI'21} & 157.0 & 30.8 & 20.50 & 1.95 \\ \hline
Multiplier-less H\&H\cite{PWL_HH_Frontiers'14} & 359.2 & 190.0 & 31.54 & 4.20 \\ \hline
RAM H\&H\cite{PWL_HH_Frontiers'14} & 317.3 & 104.0 & 35.60 & 3.85 \\ \hline
TCAD'23\cite{TCAD'23} & 18.94 & 24.35 & 6 & 1.18 \\ \hline
CORDIC Izhikevich\cite{CORDIC_Izhikevich_TBioCAS'22} & 66.0 & 17.68 & 9.29 & 1.05 \\ \hline
TCAS-I'22\cite{TCAS-I'22} & 213.0 & 352.0 & 6.68 & 2.95 \\ \hline
IF-1\cite{TVLSI25} & 102.5 & 166.7 & 11.4 & 1.365\\ \hline
LIF-1\cite{TVLSI25} & 104.1 & 169.2 & 12.7 & 1.43\\ \hline
IF-2\cite{TVLSI25} & 92.6 & 159 & 11.4 & 1.365\\ \hline
LIF-2\cite{TVLSI25} & 93.7 & 161.4 & 12.1 & 1.43\\ \hline
NC'20\cite{NC20} & 140.5 & 81.5 & 56.8 & 4.6\\ \hline
Access'22\cite{Access22} & 43.2 & 36.8 & 32.2 & 6.95\\ \hline
\textbf{Proposed} & \textbf{46.37} & \textbf{30.4} & \textbf{2.38} & \textbf{0.54}\\ \hline
\end{tabular}}
\end{table}

\section{Conclusion \& Future Work}

This work presented L-SPINE, a low-precision SIMD-enabled spiking neural compute engine for efficient edge inference. By combining multi-precision execution with a multiplier-less datapath, the proposed architecture achieves significant improvements in area, latency, and energy efficiency. The FPGA implementation demonstrates millisecond-level inference with sub-watt power consumption, achieving orders-of-magnitude improvements over conventional CPU and GPU platforms. Furthermore, quantization analysis confirms that low-bit precision enables substantial memory savings with minimal accuracy loss. These results highlight the potential of unified low-precision SIMD architectures for scalable SNN acceleration. Future work will explore layer-adaptive precision scaling for next-gen. edge AI systems.

\bibliographystyle{ieeetr}
\bibliography{bib}

\end{document}